# Multimetric Event-driven System for Long-Term Wireless Sensor Operation in SHM Application

Muhammad Zohaib Sarwar, Muhammad Rakeh Saleem, Jong-Woong Park, Do-Soo Moon and Dong Joo Kim

*Abstract*— **Wireless sensor networks (WSNs) are promising solutions for large infrastructure monitoring because of their ease of installation, computing and communication capability, and cost-effectiveness. Long-term structural health monitoring (SHM), however, is still a challenge because it requires continuous data acquisition for the detection of random events such as earthquakes and structural collapse. To achieve long-term operation, it is necessary to reduce the power consumption of sensor nodes designed to capture random events and, thus, enhance structural safety. In this paper, we present an event-based sensing system design based on an ultra-low-power microcontroller with programmable event-detection mechanism to allow continuous monitoring; the device is triggered by vibration, strain, or a timer and has a programmed threshold, resulting in ultra-low-power consumption of the sensor node. Furthermore, the proposed system can be easily reconfigured to any existing wireless sensor platform to enable ultra-low power operation. For validation, the proposed system was integrated with a commercial wireless platform to allow strain, acceleration, and time-based triggering with programmed thresholds and current consumptions of 7.43 and 0.85 mA in active and inactive modes, respectively.**

## I. INTRODUCTION

Natural disasters, such as earthquakes and typhoons, and human-induced disasters such as vehicle collisions with structures, deterioration (e.g., fatigue), or design errors (e.g., buckling) can cause unpredictable structural failure, the consequences of which can be catastrophic. In recent years, with the development of sensor technologies, wired inspection method have allowed the assessment of structural conditions in real time. However, high deployment costs, maintenance, and data management remain unresolved problems [1] [2] [3]. To overcome these challenges, structural health monitoring

This research was supported by Basic Science Research Program through the National Research Foundation of Korea (NRF) funded by the Ministry of Education(2017R1C1B5018231). (Corresponding author: Jong-Woong Park)

M. Zohaib Sarwar, M. Rakeh Saleem and Jong-Woong Park are with the Department of Civil and Environmental Eng., Chung-Ang University, Seoul, South Korea (e-mail: zohaib.sarwar@hotmail.com; rakeh@cau.ac.kr; jongwoong@cau.ac.kr )

Do-Soo Moon is with Department of Civil and Environmental Engineering, University of Hawaii at Manoa, Honolulu, Hawaii (e-mail: dsmoon@hawaii.edu ).

Dong Joo Kim is with Department of Civil and Environmental Engineering, Sejong University, Seoul, South Korea (e-mail: djkim75@sejong.ac.kr).

solutions based on wireless sensor networks have been developed, and the effectiveness of this strategy has been demonstrated by the deployment of wireless sensors in real structures [4]. Wireless sensor networks (WSN) have been widely deployed for infrastructure monitoring, e.g., bridge monitoring, earthquake monitoring, and large building monitoring, because of their ease of installation and cost-effectiveness.

A WSN comprises hundreds of nodes that communicate with each other. Each node comprises a sensing system, processing system, and communication system and has a limited power source. Battery powered WSNs have been developed to monitor and assess structural health for short and medium-term deployment, and several technical challenges facing WSNs have been addressed to enable their use in structural health monitoring (SHM) [5]. However, WSN deployment in long-term SHM is challenging because of the energy consumption of the sensor nodes [6] [7]. In long-term monitoring, it may be difficult or even impossible to change the sensor batteries. To overcome this challenge, many researchers have proposed hardware-based power management solutions [8] [9]. In an effort to reduce the power consumption of the hardware for long-term, continuous operation of the sensor node, event-triggered sensing systems have been developed. A wireless sensor platform, the TelosW platform [10], was upgraded from the TelosB platform using an ultra-low-power acceleration sensor as a wake-up mechanism with the integration of wake-on radios. Sutton *et al.,* [11] developed an event-detection system with low power wake-up receiver that used 1/1000 of the power of the designs presented in [10]. Bischoff *et al.* [12] used an event-detection-based wireless sensor system for railway bridge monitoring using a commercial accelerometer with a triggering function. Fu *et al.* [13] introduced a demand-based event monitoring system by integrating a low-power acceleration trigger sensor with a high-fidelity sensor platform, Xnode [14], and integrated two measurements to reconstruct responses before triggering so that acceleration data can be recorded during train passage. Most hardware remedies use accelerometers with triggering functions or comparators, but a multisensor triggering system has not yet been reported.

Moreover, existing triggering systems are platform dependent, thus confining them to a specific wireless sensor platform and hindering wide adoption of event-triggering systems. In SHM applications, a multisensor trigger mechanism is required that can respond to structural failure caused not only by dynamic but also, static responses such as the buckling and fatigue of steel members or concrete creep.

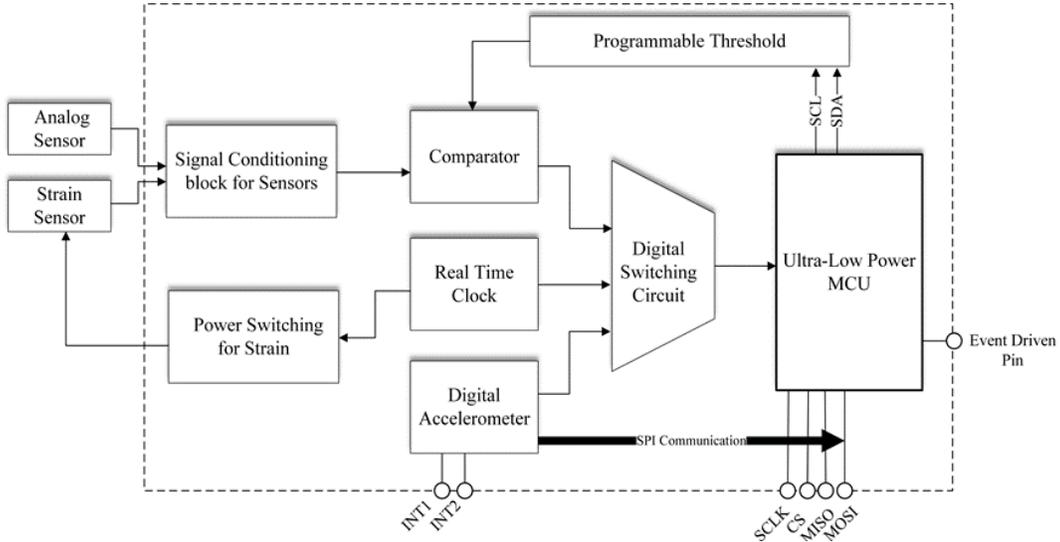

Fig. 1. Functional block diagram of the event-driven system.

In this paper, we present the design and evaluation of multimetric event-driven system (EDS) for ultra-low-power operation. The proposed system is designed based on a microcontroller with programmable event-detection logic to allow continuous monitoring and triggering by multimetric measurements such as vibration, strain, and time. Furthermore, the proposed system can be easily reconfigured with a commercial wireless sensor platform and enables ultra-low-power operation. The proposed system was tested by integrating it with a commercial wireless platform to test the triggering performance, sensitivity, and power consumption.

This paper is composed as follows: In Section 2, the proposed system model and hardware structure of the event-detection board is explained; in Section 3, we describe the integration of the multimetric EDS with a commercial high-fidelity sensing platform to validate the vibration, strain, and periodic timer triggers, followed by power analysis; and, in Section 4, we present concluding remarks and future work

## II. SYSTEM DESIGN

This section describes the design model and operating principle of the event-triggered wireless smart sensor in terms of power consumption and responsiveness. The system is divided into several modules, as shown in Fig. 1. Each module is described in the subsequent sections according to the event-triggered interface.

### A. Working Principle

The EDS generates a logic signal at a predetermined time interval in response to vibration and strain inputs that exceed a programmed threshold. In addition, a periodic wake-up signal is required for the sleep control of the sensor node to maintain the overall power of the node at the minimum level. To achieve these design goals, three basic modules were designed and implemented: (1) a vibration-based trigger, (2) a periodic wake-up trigger, and (3) a strain-based trigger. The vibration-based trigger is activated when the structure experiences a certain level of acceleration. The strain-based event trigger is designed to capture strain responses beyond the programmed limit. In this case, the strain signal is amplified, filtered, and compared to the predefined threshold to determine whether a valid event has occurred. Periodic wake-up is used to control the sleeping cycle, as well as to monitor structural health at set time intervals. All these modules work independently to generate a logic signal if any of the triggers is activated; the digital logic block and a local ultra-low-power microcontroller unit (MCU) (ATtiny85) aggregate the logic and generate an active signal to the main wireless sensor platform for high-fidelity sensing.

The overall framework of the multimetric EDS is shown in Fig. 1. The multimetric EDS is designed according to the function of each block. Design requirements and constraints for each triggering function are described in the subsequent sections.

### B. Vibration-Based Triggering

Vibration-based trigger mechanisms have been widely reported in the literature, as mentioned in Section 1. To capture a rare and sudden event, a hardware solution is required that can continuously measure the vibrations of structure and store them in its limited memory. When the vibrational responses exceed a certain threshold, the interrupt pin switches to low impedance and wakes up the whole sensing system. The power consumption of such a triggering sensor is very low, allowing it to work continuously on battery power. The other characteristic includes adequate resolution, high sensitivity, sufficient sampling rate, and a large memory buffer so that data before an event can be recovered. Several comprehensive reviews have been reported concerning different accelerometer triggers, and the ADXL362 developed by Analog Devices is considered to be the best candidate [14] [10]. The ADXL362 is three-axis microelectromechanical system (MEMS) accelerometer with a serial peripheral interface (SPI) and analog-to-digital converter (ADC). The main feature of the ADXL362 is its low-power operation; the device only consumes 13 μA in low power noise mode and has a sampling rate of 400 Hz and resolution of 1 m$g$. Its "first in, first out" buffer allows the sensors to store up to 512 data samples[15], and the sampling rate can reach 100 Hz, which corresponds to 1.7 s of data. Further, the ADXL362 has built-in logic for activity detection on the basis of an acceleration threshold, which can be used as a triggering mechanism. The vibration-based triggering is implemented using the ADXL362 through the SPI controlled by a local MCU (Attiny85), where the parameter for the accelerometer is set for activity detection. The software framework for the triggering mechanism will be covered in Section III C.

### C.   *Strain-Based Triggering Interface:*

For strain-based triggering, the design of the amplifier is important to ensure the measurement of strain response within a few 100 μ$\varepsilon$, which is generally required in SHM. A Wheatstone bridge was used to measure the strain. Let us first consider the relationship between the Wheatstone bridge output voltage and strain, as given by (1):

$$\Delta V = \frac{G.F \cdot \varepsilon}{4} V_{EXT} = \frac{\Delta R}{4R} V_{EXT} , \qquad (1)$$

where $\varepsilon$ is given by (2).

$$\varepsilon = \frac{\Delta V \times 4}{G.F \times V_{EXT}} \qquad (2)$$

$G.F$ is the gauge factor of the strain gauge, $V_{EXT}$ is the input voltage, and $R$ is the resistance of the Wheatstone bridge circuit. The output voltage for a strain of 100 μ$\varepsilon$ calculated with (1) is 0.16 mV with a gauge factor of 2 and an input voltage of 3.2 V. The output voltage is so small that even a change in 100 μ$\varepsilon$ may not be detected by the comparator circuit. Thus, signal conditioning for amplification and filtering of $\Delta V$ is required to enable precise strain triggering. Fig. 2 shows a schematic diagram of the strain-based trigger comprising a Wheatstone bridge, instrument amplifier, low-pass filter, and comparator.

The Wheatstone bridge output voltage, $\Delta V$ , is amplified by a precision instrument amplifier, an INA2128, which supports a gain factor of up to 10000 times and has a low quiescent current with 700 μA[16]. The gain of the INA2128 is programmed by connecting the single external resistor, which is given by

$$Gain = 1 + \frac{50k\Omega}{R_G} , \qquad (3)$$

where $R_G$ is a gain-setting resistor in the design of the strain-based trigger, a resistance of 470 Ω was implemented, and the corresponding gain is 107.4. Because the amplifier works only in a positive supply range, a voltage offset of 1 V is added to the amplifier output, given as

$$V_o = Gain(\Delta V) + V_{offset} . \qquad (4)$$

The amplified signal, $V_o$, is then low-pass-filtered to eliminate undesired amplified signal noise and fed into the comparator to be compared with the programmed threshold. The low-pass filter was designed with a first-order passive resistor–capacitor

(RC) filter with resistance and capacitance of 1 kΩ and 10 μF, respectively, to yield a cut-off frequency at 16 Hz. Because the responsiveness of the comparator module depends upon comparator propagation delay and filter time constant ($\tau = RC$), the small values of R and C are used to lower the response time.

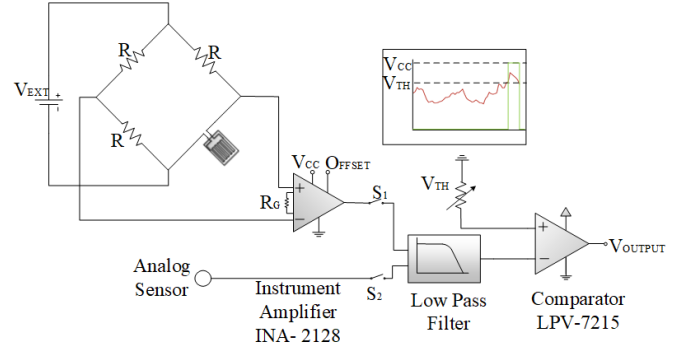

Fig. 2. Functional schematic of strain-based triggering.

The comparator circuit was designed using a LPV7215 from Texas Instruments [17, p. 72]. The main feature of this comparator is that it has very low propagation delay, typically 4.5 μs, as well as a low current consumption of 580 nA. The comparator compares, first, two analog inputs from the amplified and low-pass-filtered strain signal and, secondly, the programmed threshold voltage, generating an output logic signal of 1 when the strain signal is greater than the programmed threshold and logic signal of 0 otherwise.

For the accurate programming of the threshold voltage, a voltage divider with a non-volatile digital potentiometer, a MAX5479, was used to set up the threshold voltage ($V_{TH}$) as

$$V_{TH} = \frac{R_{DP}}{100,000} V_{EXT} , \qquad (5)$$

where $R_{DP}$ is the resistance of the MAX5479, and $V_{EXT}$ is input voltage of 3.2 V. The MAX5479 is a non-volatile digital potentiometer having a resistance value of 100 kΩ with 256 tap positions [18, p. 54], so that $R_{DP}$ can be set with 256 steps. The minimum change of strain threshold that can be programmed is

$$V_{TH\_MIN} = Gain \times \Delta V , \qquad (6)$$

where $V_{TH\_MIN} = V_{EXT}/256$. Combining (6) and (1), the minimum programmable strain for triggering is calculated as

$$\varepsilon_{TH\_MIN} = \frac{4}{256 \times GF \times Gain} \qquad (7)$$

The minimum value of $\varepsilon_{TH}$ is 70.3 μ$\varepsilon$ with a gain of 107.4 and *G.F.* of 2.07.

The total current consumption of the strain-based triggering module in a quiescent state (when no event occurs) is 120 μA and additional current consumption for the strain gauge is calculated as $V_{EXT}/2R = 4.57$ mA. The power consumption arising from the gauge connection was reduced by designing the strain power switch to control the power in the Wheatstone bridge selectively. Power switching for strain is explained in Subsection E.

### D. *Real Time Clock*

Traditionally, a wireless sensor node uses duty-cycle and demand-based methods. To retain that functionality while using event trigger-based mechanism, hardware must be considered. To achieve this goal, a real-time clock (RTC) was employed. The RTC generate logic signals from an alarm at user-specified times. The proposed EDS uses a DS1342 as an RTC because of its low current consumption of 250 nA with equivalent series resistance (ESR) crystal compatibility [19]. The DS1342 can save the date of any event or abnormality when the EDS is triggered and provides two pins for alarm control that generate two alarms to drive logic signals at predefined time slots. The two pins for alarms are defined as Alarm1 and Alarm2; Alarm1 can be set to provide timer-based triggering and Alarm2 is programmed to control the duty cycle for the strain power switch.

### E. *Strain Power Switch*

Strain sensing is one of the most power-consuming parts of the proposed multimetric EDS because one active stain gauge of $350\,\Omega$ can consume $V_{EXT}/2R = 4.57$ mA, and the instrument amplifier consumes 1.4 mA with a $V_{EXT}$ of 3.2 V; the total power consumption required for strain sensing is 5.97 mA. The power consumption can be minimized by introducing a user-defined duty cycle to control the power of strain gauge for a predefined time such that strain response is stabilized and reliably triggered.

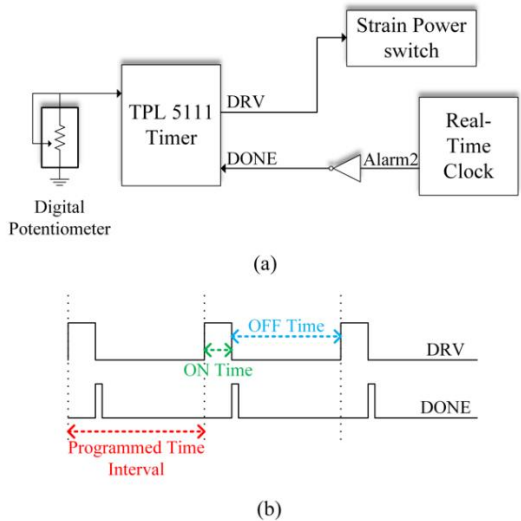

(a)

(b)

Fig. 3. (a) Power switching for strain and (b) timing diagram.

To control the power for strain sensing, an RTC and a TPL5111 nanopower system timer were used (see Fig. 3). The TPL5111 is designed to control the duty cycle of the timer based on the external resistance set by a digital potentiometer [20] [21]. The TPL5111 timer produces periodic pulses through the DRV pin at a certain time interval according to the programmed timing. For example, an external resistor of 125 kΩ yields a 1 h time interval. The length of the pulse in the

DRV pin denoted as 'On Time' in Fig. 3 is set by receiving a DONE pulse from the RTC. The logic signal from the DRV pin is used to control the power of the Wheatstone bridge. For example, if 125 kΩ is attached to the TPL5111 and Alarm1 is set to fire once every hour, the DRV is generated every hour with a length of 1 min.

### F. *Digital Switches and Microcontroller Unit*

The logic signals from different trigger modules must be aggregated before being fed to the MCU so that the MCU can decide whether to switch on power and wake up the main sensor platform. For signal processing, an OR logic circuit was implemented in the hardware and the MCU is controlled via software programming.

The main component of the OR logic circuit is an SN74AUP low-power single buffer/driver that typically consumes 0.9 μA. When any of the triggering modules generate a logically high signal, the OR logic circuit instantly passes the logic to the MCU; the logic signal is held for predetermined time interval until the main sensor platform wakes up.

The MCU considered in the study is a Microchip ATtiny85, 8 bit-AVR [22]. The ATtiny85 has program memory of 8 KB and digital communication peripherals (SPI, I2C). Because of its low power consumption and rich digital interface, the ATtiny85 is used for controlling the digital potentiometer, RTC, and ADXL362, as well as to hold the OR logic.

The ATtiny85 has an internal clock range from 1 to 20 MHz and two different power modes were designed: (1) power-down mode and (2) active mode. In power-down mode, all ATtiny peripherals are powered off except external pin change interrupt. In that mode, the ATtiny85 consumes 0.1 μA at 1 MHz. The initialization time from power-down mode to active mode by pin change interrupt is within two cycles of clock frequency, which is sufficient to capture dynamic signals. In active mode, the MCU consumes 300 μA at 1 MHz; the active mode is used for holding the logic signal until it is received by the main sensor platform.

## III. System Integration and Evaluation

In this section, integration of the system design with a state-of-the-art high-fidelity sensing platform is discussed. The section includes a brief overview of the high-fidelity sensing platform, Xnode. Software and hardware consideration for the integration of the proposed multimetric EDS with the Xnode is discussed. In addition, a laboratory-scale experiment was carried out to demonstrate the performance of the proposed multimetric EDS in terms of responsiveness and power consumption.

### A. *Integration with a High-Fidelity Sensing Platform*

The requirements of a wireless sensor for SHM are a powerful processing unit that can handle a large amount of data and a data acquisition system that can obtain precise data from sensors at different sampling rates. In this study, we adopted a commercial Xnode wireless sensor because of its powerful computational ability and high-precision data acquisition suited for SHM. The Xnode platform consists of three boards: (1) an MCU board that is an off-the-shelf commercial MINI4357 board manufactured by Embest [23], (2) a radio board that controls the power of the system and radio frequency communication, and (3) a sensor board that has an 8-channel, 24-bit ADC (i.e., ADS131E08 [24]), three-axis analog accelerometer (i.e., LIS344ALH [25]), and a Wheatstone bridge for strain sensing. On the software side, the Xnode has open-source middleware services and it supports custom application development [26].

The multimetric EDS is integrated with Xnode by connecting the output logic signal from the multimetric EDS to the digital power switch in the radio board, thus controlling the whole power supply on the system by digital logic.

### B. Printed Circuit Board Design of the Multimetric EDS

The printed circuit board (PCB) layout of the proposed multimetric EDS is designed to have stackable radio and sensor boards to allow integration in a single sensor unit (see Fig. 3). Fig. 3(a) shows the important components, including the MCU, RTC, OR logic block, and strain amplifier. The component denoted as a debugging port is a six-pin connector for programming the ATtiny85 MCU. To interface with the Xnode, 80-pin connectors were used for sharing power, ground, and general-purpose input/output. The hardware integration between the Multimetric EDS and the Xnode is illustrated in Fig. 4.

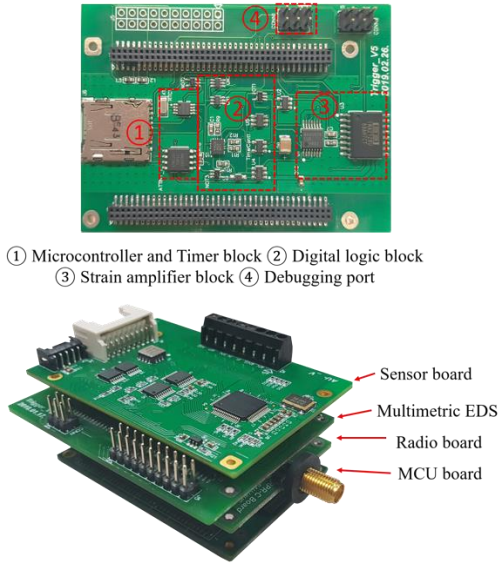

① Microcontroller and Timer block ② Digital logic block
③ Strain amplifier block ④ Debugging port

Sensor board
Multimetric EDS
Radio board
MCU board

Fig. 4. PCB implementation of EDS: Integration of Xnode platform and EDS.

Because the sensor board has a Wheatstone bridge and digital power switch for strain power, the multimetric EDS only drives logic signals to control the power on the strain circuit for certain intervals and durations as determined by the strain power switching (see Section II D). The multimetric EDS receives differential strain signals from the sensor board for event detection. The radio board has two power sources: one power rail is dedicated to supplying a constant 3.2 V to the multimetric EDS, and the other rail supplies power to the whole sensor node activated by digital logic in the radio board that is controlled by the multimetric EDS. Furthermore, the digital logic signals from each triggering module are linked to the MCU board to indicate the type of event.

### C. Software Framework for Integration

In addition to hardware integration, a software application framework was developed to implement and integrate the multimetric EDS with the Xnode.

Fig. 5 shows the framework and flowchart of the event triggering and sensing application. The flowchart is divided into three phases: (1) startup initialization, (2) event detection, and (3) high-fidelity sensing through the Xnode. When power is on for the Multimetric EDS, the ATtiny85 MCU in the EDS loads all parameters, including threshold values for strain, configuration setup for activity detection threshold for ADXL362, and Alarm2 on RTC for strain power switching and Alarm1 for the timer-based trigger. After initialization, the multimetric EDS goes into power-down mode and waits for an interrupt.

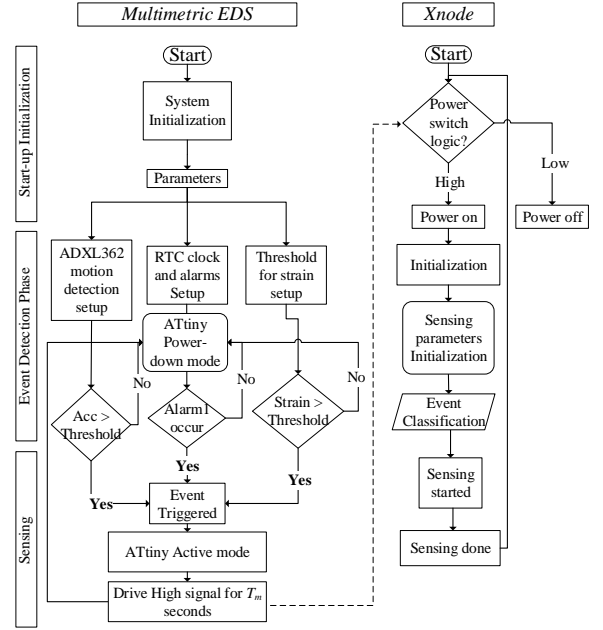

Fig. 5. Flowchart of event-triggered sensing.

In power-down mode, the ATtiny85 turns off all functions except the external pin interrupt that waits for a trigger to generate a digital high logic signal. If any of the thresholds are exceeded (i.e., an event occurs), the ATtiny85 switches into active mode and drives the digital high logic for $T_m$ to the radio board to power up the Xnode. For power saving, $T_m$ is adjustable and is set 30 ms. On startup of the Xnode, a latching switch for power is set to maintain its state until the sensing task is complete. The sensing parameter is initialized, and the time of the captured event and the type of event is saved in the NAND flash memory. The type of event is classified based on the digital logic generated by the trigger modules. The sensing task is implemented, and the sensing data is saved in the SD memory. After the sensing task, the Xnode releases the latch switch to turn off the system.

### D. *Evaluation of the EDS*

Experimental validation of the multimetric EDS integrated with Xnode was carried out through laboratory-scale tests using an aluminum cantilever beam, as shown in Fig. 6. The length and width of the beam were 1000 and 100 mm, respectively. The proposed system was installed at the top of the beam to measure the acceleration at the top, and the strain gauge was attached to the bottom of the beam because acceleration and strain are maximized at the top and the support of the cantilever structure, respectively.

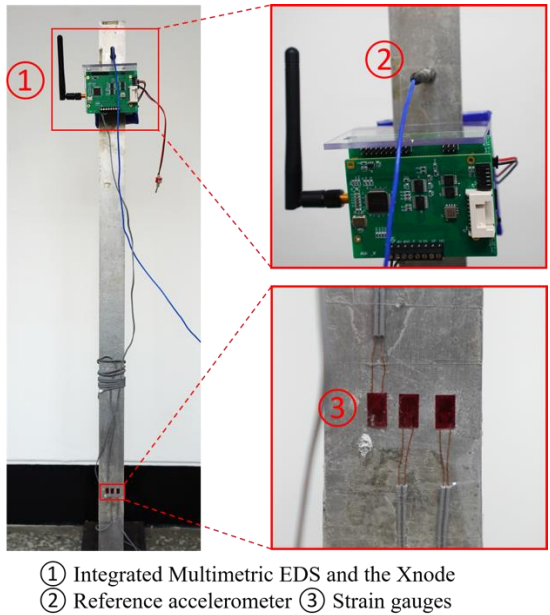

① Integrated Multimetric EDS and the Xnode
② Reference accelerometer ③ Strain gauges

Fig. 6. Experimental EDS setup.

The sampling rate of the proposed system was set to 1000 Hz. As a reference measurement, a PCB353B33 wired accelerometer [27] was installed at the same location as the proposed system, and the reference strain with a Wheatstone bridge was measured using a LabJack U6 Pro [28] with a data sampling rate of 1000 Hz. Two experimental scenarios were designed to evaluate vibration-based and strain-based triggers.

The vibration trigger was first evaluated based on a threshold of 80 m$g$. The ADXL362 in the multimetric EDS was set to operate in an ultra-low-noise mode measuring vibration at 100 Hz for accurate event detection. An impact force was applied at the top of the beam, and the multimetric EDS was triggered at 9.5 s when the first peak vibration crossed the threshold of 80 m$g$. The time between the actual event detection and sensing by the Xnode was 0.95 s, most of which was the Xnode startup. After the Xnode began sensing, both acceleration events from the reference and the Xnode showed good agreement, validating the vibration-based triggering system.

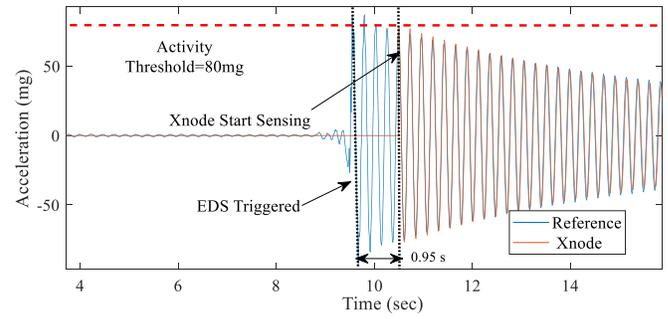

Fig. 7. Evaluation of vibration trigger

The strain-based triggering was validated by applying increasing static force at the top of the beam to trigger only the strain response and minimize vibrations. The strain-based triggering was validated using a threshold value of $\varepsilon_{TH}$ =1160 με. Note that the multimetric EDS system conducts shunt calibration to obtain the scaling factor between the change in the voltage and the actual strain response, and the voltage for the threshold is determined for threshold setting.

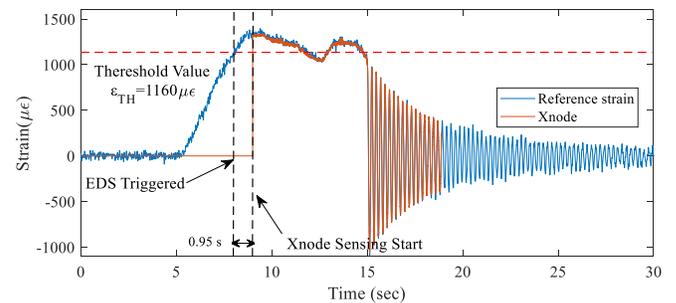

Fig. 8. Evaluation of strain trigger

The applied strain was gradually increased and, when it exceeded the threshold strain, $\varepsilon_{TH}$, at 8.132 s, the multimetric EDS was triggered and started sensing at 9.05 s. The time between the actual event detection and sensing was 0.95 s, the same as the vibration-based triggering system. Figs. 7 and 8 demonstrate the ability of the event-driven system to detect events caused by vibration or strain efficiently for rapid structural condition assessment.

In addition to the evaluation of the vibration and strain-based triggers, daily event triggering was evaluated. A similar experimental setup to that shown in Fig. 6 was used. For the experiment, the RTC was configured for a twice daily alarm at 7:05 am and 7:05 pm. The threshold value for acceleration and strain were set as 200 m$g$ and 264 με, respectively. During the experiment, the beam was excited by impact and static forces at the top of beam at random times throughout the day. The purpose of the experiment was to evaluate the performance of the multimetric sensor and time-based event classification. The total number of events recorded over a period of one day was 20. In total, 13 events were triggered by vibration, and 5 events were triggered when the threshold strain value was exceeded.

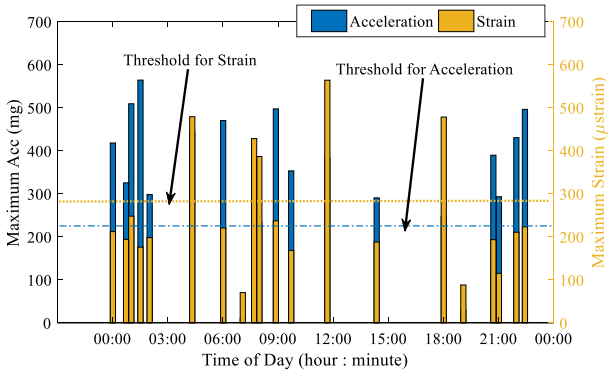

Fig. 9. Long-term EDS validation.

As shown in Fig. 9, at 1:01 am, an event was recorded, and the maximum acceleration of the vibration during the event was 509.5 m*g*, which is well above the acceleration threshold and the maximum strain was 248.5 με Thus, the event is classified as an acceleration trigger. Similarly, at 11:41 am, the sensor node was triggered by the strain sensor because the maximum value of strain at this time was above the threshold value. Two events were recorded during the experiment when the maximum values of strain and acceleration were less than a predefined threshold. Both events correspond to time-based triggering, and the recorded times of the events were 7:05 am and 7:05 pm.

### E. *Current Consumption Analysis for the Multimetric EDS*

The current consumption of the multimetric EDS is listed in Table I.

TABLE I: Current consumption of the multimetric EDS

| Module | Inactive Mode | Active Mode | Activity Time |
|---|---|---|---|
| **Comparator** | 580 nA | 580 nA | Always ON |
| **Instrument Amplifier** | 450 μA to 1.4 mA[a] | 1.4 mA[a] | User-Defined Time |
| **Wheatstone bridge** | 0–4.57 mA[a] | 4.57 mA | |
| **Potentiometer** | 300 μA | 300 μA | Always ON |
| **Timer** | 35 nA | 35 nA | Always ON |
| **Digital Logic Block** | 0.9 μA | 1.6 μA | 5.5 ns/Event |
| **ATtiny85** | 0.1 μA | 0.9 mA | 4 ms/Event |
| **Real Time Clock** | 250 nA | 250 nA | 20 ms/Event/Alarm |
| **ADXL362** | 13 μA | 13 μA | Always ON |
| **Total** | 764 μA to 6.28 mA | 7.43 mA | 35 ms/Event [b] |

[a] Strain power is controlled by a programmable timer and can be triggered at a predefined time to maintain the overall current at the minimum level in the active mode.
[b] Activity time corresponds to the time required for the multimetric EDS to drive the logic signal to the main sensor platform

The multimetric EDS has two states of operation: inactive and active. The multimetric EDS is continuously measuring responses with the power-down mode of the ATtiny85 in inactive mode consuming 764 μA when the strain power switch is turned off and 6.28 mA when strain power is turned on. In active mode, all modules are working and the ATtiny85 is switched to active mode and drives logic signals to the main sensor platform, in the process consuming 7.43 mA. Note that the activity time for active mode is very short, 30 ms, so that the current consumption in active mode is negligible.

Assuming that the period of strain power switching can be set to 1 min every hour, the average current consumption of the multimetric EDS is calculated as 856 μA (i.e., 1-min power consumption @ 764 μA and 59-min power consumption @ 6.28 mA).

### F. *Power Consumption of the Integrated System*

The power consumption of the integrated system of the multimetric EDS and the Xnode was investigated. The average current drawn by the Xnode [14] in the sensing stage is 170 mA. Using a 3.7-V DC lithium–polymer battery of 10000 mAh, the Xnode only lasts for three days [13].

The service life of the integrated system depends on the number of events, which can be described by probability of an event, $P_D$ (i.e., the amount of time for sensing triggered by events over 100 h). The service life can be calculated as

$$Service_{life} = \frac{Capacity(mAh)}{I_{AVG}} \times 0.8 \qquad (8)$$

where 0.8 is a compensation factor for an external and environmental factor that affect battery life [29]. The average current consumption during operation can be expressed using (9).

$$I_{AVG} = I_{IDLE}(1 - T_A) + I_{SENSING}(T_A) \qquad (9)$$

Here, $T_A$ is the sensing duration using the Xnode when events are detected, $I_{IDLE}$ is 856 μA, and $I_{SENSING}$ is 172.86 mA; the average current for the sensing task is the sum of the current consumption of the Xnode [13] of 170 mA and multimetric EDS in active mode of 7.86 mA (see Table I). Assuming 1% $P_D$, indicating 10 events per day with 2 min sensing per event, the $I_{AVG}$ current is calculated as 2.62 mA with a 10 Ah battery using (9), and the corresponding service life is 3054 h. The service life and average current consumption are summarized according to different $P_D$ values in Table II.

TABLE II: Service life for the integrated system

| Probability of Event ($P_D$) | Average Current ($I_{AVG}$, mA) | Service Life (h) |
|---|---|---|
| Case 1: 1% | 2.62 | 3054 |
| Case 2: 5% | 9.68 | 826 |
| Case 3: 10% | 18.51 | 432 |
| Case 4: 20% | 36.17 | 221 |

Thus, the Xnode integrated with the multimetric EDS has an extended service life of up to 3054 h; this makes it suitable for continuous vibration and strain monitoring in SHM.

### G. Comparison with state of art system

This section summarizes the comparison with state of art solution available for applications of high-fidelity sensing in structural health monitoring. Table III shows the comparison based on triggering mechanism and always ON sensing power consumption.

TABLE III: Comparison with state of art solutions in SHM

| Sensing Node | Triggering Mechanism | Always ON Sensing |
|---|---|---|
| Narada [30] | Duty Cycle Only | ~300mW |
| Imote 2+SHMA [31] | Duty Cycle Only | ~400mW |
| Martlet [32] | Duty Cycle Only | ~450mW |
| Waspmote [33] | Acceleration Only | N/A |
| Xnode [14] | Duty Cycle Only | ~630mW |
| Integrated EDS | Multimetric Sensors | ~24mW-~630mW [a] |

[a] Integrated EDS module continuously remain ON and do the sensing in low power mode and in case of triggering event, it turn ON the high-fidelity sensing platform to record and process the data.

From comparison and lab scale experiments integrated EDS had shown potential to be used in long term monitoring with minimum power budget in comparison to state of art solutions (see Table III). The main advantage of proposed EDS module is it can easily be reconfigured and integrated with above mention solutions without compromising the performance of the sensor node.

## IV. CONCLUSION

This paper proposes an event detection and triggering system for the continuous and long-term operation of a wireless sensor for the monitoring of civil infrastructure that is vulnerable to rare events (i.e., natural disasters and progressive failure). The proposed multimetric EDS developed in this study enables the following:

- Multimetric event detection for the continuous monitoring of structures in response to random and critical events.
- Long-term operation by design of power-switching circuits and the use of an ultra-low-power MCU.
- Reconfigurability and integrability with commercial wireless sensor platforms to enable ultra-low-power operation.

To achieve these results, three basic triggering modules (vibration, strain, and time-based triggers) were designed with an ultra-low-power ICS; these modules can be programmed to adjust the threshold of response and timing for event detection. All these modules work independently to generate logic signals if any of the triggers is activated; the digital logic block and an ultra-low-power MCU (an ATtiny85) aggregate the logic and wake up the main wireless sensor platform for high-fidelity sensing. Moreover, power switching for high-power-consumption components, such as the instrument amplifier in the strain trigger, can be controlled by the MCU.

The developed multimetric EDS was integrated with a commercial wireless sensor platform, Xnode, for validation. The integrated system was experimentally validated on a cantilever beam with inputs from impact and static forces used to activate the vibration and strain-based triggers, respectively. The results indicate that both triggering modules are activated at the predefined thresholds and show good agreement with the results of reference accelerometer and strain sensors. The time between the detection of an event by the multimetric EDS and the start of sensing by the Xnode was calculated to be 0.95 s, which is sufficient for rare event detection. In addition, performance evaluation over the course of one day was conducted with the same experimental setup. A total of 20 events were recorded, 13 of which were triggered by vibration and 5 of which were triggered by strain. Two additional events were triggered with the programmed timer.

Power consumption analysis of the multimetric EDS and the integrated system was also carried out, and the average current consumption of the multimetric EDS was found to be 856 µA considering periodic power switching for the instrumental amplifier (i.e., 1 min every hour). In addition, under the assumption of a 1% event probability, the integrated system is expected to last 3054 h with a 10 Ah battery, representing a significant extension from the three days without the multimetric EDS.

In summary, the developed multimetric EDS is promising for long-term SHM applications. For the timely detection of random events and rapid decision making, the activation of the sensor network and the collection synchronized measurements for structural condition assessment is crucial. Future work will include the integration of a low-power event trigger antenna for waking the whole wireless sensor network in response to a certain event and a time synchronization strategy among the different triggering nodes.

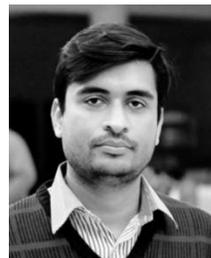

**M. Zohaib Sarwar** received a B.S. degree in electrical engineering from Bahria University, Pakistan, 2017. He is currently working toward his Master's degree in civil engineering, majoring in structural engineering at the Smart Infrastructure Laboratory, Chung-Ang University, Seoul, Republic of Korea. His current research interests include structural health monitoring, wireless smart sensor networks, and sensor fusion for structural assessment.

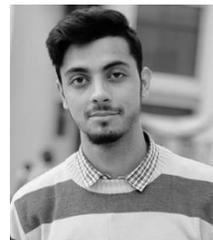

**M. Rakeh Saleem** received a B.S. degree in electrical engineering from Bahria University, Islamabad, Pakistan in 2017. He is currently pursuing a Master's degree in structural engineering at Chung-Ang University, Seoul, Republic of Korea. Until May 2018, he worked as a development engineer at the National University of Science and Technology, Islamabad, Pakistan for one year. His research interest includes machine learning, wireless sensor networks, and structural health monitoring.

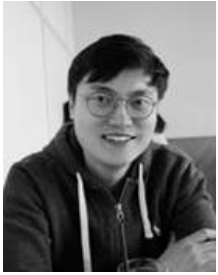 **Jong-Woong Park** received a B.S. degree in civil engineering from Hanyang University in Korea in 2008, and M.S. and Ph.D. degrees in civil engineering from KAIST in 2009 and 2013, respectively. He worked as a postdoctoral researcher at the University of Illinois at Urbana-Champaign in the U.S.A. from 2014 to 2017. He currently works as an assistant professor in the Department of Civil and Environmental Engineering at Chung-Ang University, Seoul, Republic of Korea. His current research interests are structural health monitoring using wireless smart sensors, image processing for structural response analysis, and structural dynamics.

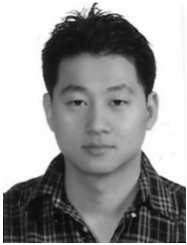 **Do-Soo Moon** obtained his B.S. and M.S. degrees in civil engineering from Seoul National University in Korea in 2001 and 2006, respectively, and Ph.D. degree in civil engineering from University of Illinois at Urbana-Champaign in the U.S.A. in 2012. He is currently working as an assistant professor in the Department of Civil and Environmental Engineering at the University of Hawaii at Manoa. His research interests include: multi-hazard sustainable and resilient structures for civil infrastructures; risk assessment and mitigation; structural health monitoring with smart sensors; evaluation of structural performances under extreme events; and information technology and structural control applications in structural engineering.

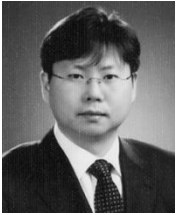 **Dong Joo Kim** received his B.S. and M.S. degrees in civil engineering from Hanyang University in Korea in 1997 and 1999, respectively, and Ph.D. degree in civil and environmental engineering from The University of Michigan, Ann Arbor, MI, U.S.A. in 2009. He is currently working as a professor in the Department of Civil and Environmental Engineering at SEJONG University, Seoul, Korea. His research interests include smart construction materials with self-damage and/or stress sensing capability in addition to the development of high-performance construction materials with high impact resistance.